\documentclass[aps,twocolumn,preprintnumbers,prb,superscriptaddress]{revtex4}

\usepackage{amssymb}
\usepackage{graphicx}
\usepackage{amsmath}

\begin{document}

\title{A Density Difference Based Analysis of Orbital--Dependent Exchange--Correlation Functionals}
\author{I. Grabowski}
\affiliation{Institute of Physics, Faculty of Physics, Astronomy and Informatics, Nicolaus Copernicus University, Grudziadzka 5, 87-100 Torun, Poland}
\author{A. M. Teale}
\affiliation{School of Chemistry, University of Nottingham, University Park, Nottingham, NG7 2RD, UK.}
\author{E. Fabiano}
\affiliation{National Nanotechnology Laboratory, Istituto Nanoscienze--CNR, Via per Arnesano, I-73100 Lecce, Italy}
\author{S. \'Smiga}
\affiliation{Institute of Physics, Faculty of Physics, Astronomy and Informatics, Nicolaus Copernicus University, Grudziadzka 5, 87-100 Torun, Poland}
\author{A. Buksztel}
\affiliation{Institute of Physics, Faculty of Physics, Astronomy and Informatics, Nicolaus Copernicus University, Grudziadzka 5, 87-100 Torun, Poland}
\author{F. Della Sala}
\affiliation{National Nanotechnology Laboratory, Istituto Nanoscienze--CNR, Via per Arnesano, I-73100 Lecce, Italy}
\affiliation{Istituto Italiano di Tecnologia (IIT), Center for Biomolecular Nanotechnologies, Via Barsanti, I-73010 Arnesano, Italy}

\begin{abstract}
We present a density difference based analysis for a range of orbital--dependent
Kohn--Sham functionals. Results 
for atoms, some members of the neon isoelectronic series and small molecules are reported
and compared with ab initio wave-function calculations.
Particular attention is paid to the quality of approximations to the 
exchange--only optimized effective potential (OEP) approach: we consider both the Localized 
Hartree Fock as well as the Krieger-Li-Iafrate
methods.
Analysis of density differences at the
exchange--only level reveals the impact the approximations have on the resulting electronic densities. 
These differences 
are further quantified in terms of the ground state energies, frontier orbital energy differences and 
highest occupied orbital energies obtained. 
At the correlated level an OEP approach based on a perturbative
second--order correlation energy expression is shown to deliver results comparable with those from 
traditional wave function approaches, making it suitable for use as a benchmark against which to compare standard
density--functional approximations.  
\end{abstract}

\maketitle

\section{Introduction}
The accuracy of density functional theory (DFT) within the Kohn--Sham (KS)
approach (KS--DFT) \cite{kohn:1965:KS} is strongly dependent on the approximations used in practical
exchange--correlation (XC) functionals. Although formally an exact theory,
 based on the Hohenberg--Kohn theorems \cite{hohenberg:1964:HK}, after four decades 
of growing applications and success, it still struggles with the problem of
defining theoretically correct, non--empirical, robust and practically applicable
XC functionals. 
In the last decade, significant attention has been given to
the use of orbital--dependent  XC  functionals in the KS methodology at both the
exchange 
\cite{krieger:1990:OEP,engel:1993:OEP1,grabo:1999:OEPthick,gorling:1999:OEP,ivanov:1999:OEP,hirata:2001:OEPU,lhf1,ceda,lhfopen,elp,heaton:2007:sn,Hessel1}
 and correlation 
\cite{gorling:1994:OEP,grabowski:2002:OEPP2,bartlett:2005:abinit2,grabowski:2005:1shot,mori-sanchez:2005:oeppt2,bru05,engel:2005:oeppt2,schweigert:2006:pt2,denis:2006:openshell,grabowski:2007:ccpt2,ls2,hellgren07,grabowski:2008:ijqc,weim08,hell10,verma:044105,hellgren12,oeprpascf13,sos_oep2:prb}
 levels, which  open up new routes in the search 
for new DFT methods. 

The use of orbital--dependent functionals naturally leads to the 
optimized effective potential (OEP) method~\cite{sharp:1953:OEP,talman:1976:OEP},
which defines local  
KS potentials~\cite{krieger:1990:OEP,gorling:1995:IJQCS,gorling:1994:OEP,grabowski:2002:OEPP2}
(for a review see \cite{kummel:2008:oep,lhf2,engelrev}).
To develop orbital--dependent functionals experience from wave function theories
(WFTs) can be exploited to define a series of XC approximations that systematically converge towards 
the full configuration interaction (FCI) limit~\cite{bartlett:2010:molphysOEP,grabowski:2010:molphys,grabowski:2011:jcp}.
This concept has been named {\it ab initio} DFT
\cite{bartlett:2005:abinit2}  and has proven to be a very
effective route for defining and  deriving orbital--dependent
exchange--correlation  functionals
and  potentials. Recent applications of \textit{ab initio} DFT
\cite{bartlett:2005:abinit2,grabowski:2002:OEPP2,grabowski:2007:ccpt2}
show that these functionals are free of many of the limitations of standard DFT.
There is no self interaction error problem, they provide qualitatively
correct exchange--correlation  potentials, total and correlation energies and  ionization potentials.
They have also been successfully applied to the description of van der Waals 
interactions~\cite{lotrich:2005:vdw} and systems with quasidegeneracy~\cite{grabowski:2007:ccpt2,grabowski:2008:ijqc}.

Recently the concept of
difference radial--density (DRD) distributions,
defined with respect to the Hartree--Fock (HF) radial density \cite{Jankowski:2009:DRD},
has been used to compare the electronic densities calculated from DFT approaches (both standard
and orbital--dependent) and WFT methods. In fact, it has been shown that 
the DRD distribution
\begin{equation}\label{drd_eq}
\mathrm{DRD}(r) = 4\pi r^2\left[\rho(r)-\rho^{\text{HF}}(r)\right]\ ,
\end{equation}
with $\rho$ any DFT or WFT  density and $\rho^{\text{HF}}$ the Hartree--Fock
density,
can provide a useful tool in the development and testing of new and existing
exchange--correlation functionals in KS-DFT. Based mainly on the DRD analysis it was
shown~\cite{Jankowski:2009:DRD,Jankowski:2010:DRD,grabowski:2011:jcp} that 
VWN5 \cite{vosko:1980:VWN}, LYP \cite{lee:1988:BLYP} and other correlation functionals do not individually represent  
substantial dynamic correlation effects in the KS potential or electron
density. Additionally, we have demonstrated that at the exchange--only level of approximation, popular 
standard DFT exchange functionals, in addition to
their nominal role, represent some dynamic correlation effects~\cite{Jankowski:2009:DRD,Jankowski:2010:DRD}.

In this paper we consider this analysis further and apply it to a range of
orbital--dependent exchange--only and exchange--correlation approximations. At
the exchange--only level we consider electronic densities calculated using
the Localized HF (LHF)~\cite{lhf1} and
Krieger--Li--Iafrate (KLI)~\cite{krieger:1990:OEP}  approximations. The DRDs
constructed for these approaches are compared with those from exchange--only
OEP as well as correlated {\it ab initio} DFT and WFT methods. The impact of
the approximations involved in the LHF and KLI approaches on the DRDs are
analysed in light of these comparisons. Additionally, the quality of each approach is assessed 
in terms of its associated total energy and the accuracy of the differences of
highest occupied molecular orbital (HOMO) energy and lowest unoccupied
molecular orbital energy (LUMO), the HOMO--LUMO gap. The latter provides a sensitive probe of the quality of the underlying KS effective potential. 
Accurate data for the HOMO--LUMO gaps are obtained by employing an inversion approach~\cite{wu:2003:wy} using
coupled--cluster singles--doubles and perturbative triples [CCSD(T)] densities.

\section{Methodology}
In order to define orbital--dependent XC functionals, potentials 
and correlated OEP--KS equations we
will follow the idea of {\it ab initio} DFT \cite{bartlett:2005:abinit2},
 in which the
density condition \cite{gorling:1995:IJQCS} together with coupled--cluster
(CC) methodology is employed 
to derive orbital--dependent multiplicative
exchange--correlation  potentials
in the KS--OEP approach, defining a range of exchange--only and correlated OEP methods. 

The KS density condition~\cite{ShamSchluter, gorling:1995:IJQCS}
 states that, since by construction the KS determinant
 $\Phi_\mathrm{KS}$
provides the exact density at a given space--spin coordinate,
any corrections to the converged KS density, introduced by
 changes in $\varphi_i(\mathbf{r})$, must
vanish~\cite{gorling:1995:IJQCS,bartlett:2005:abinit2},
\begin{align}\label{Eq:dencon}
\rho(\mathbf{r}) = \rho^{\mathrm{KS}}(\mathbf{r}) +
\delta \rho^{\mathrm{KS}}(\mathbf{r}), \hspace{0.5in} \delta
\rho^{\mathrm{KS}}(\mathbf{r}) = 0.
\end{align}
 The density corrections $\delta \rho^{\mathrm{KS}}(\mathbf{r})$ are  
written using the density matrix
correction $\Delta \gamma_{pq}$ from  CC
theory~\cite{bartlett:1995:coupled}  or many--body perturbation theory
(MBPT), and then can be  expanded by
orders in the perturbation $V$ by separating the total Hamiltonian,
\begin{align}
H = \sum_i h(i) + \sum_{i<j} 1/r_{ij}
\end{align}
into a zeroth--order part $H_0$ and a perturbation $V$,
\begin{align}
H = H_0 + V, \hspace{0.5in} V = H - H_0
\end{align}
For the detailed diagrammatic and algebraic derivation of working correlated
OEP equations see Refs.
\onlinecite{bartlett:2005:abinit2,grabowski:2005:1shot,schweigert:2006:pt2,grabowski:2007:ccpt2}.
\par The density condition  at first order, $\delta \rho^{(1)}(\mathbf{r})=0$,
leads to
 the exchange--only OEP (OEPx) equation,
\begin{align}
\sum_{ai} \varphi_a(\mathbf{r}) \varphi_i^*(\mathbf{r}) \left[
\frac{\langle a | \hat{K} +v_\mathrm{x} | i \rangle}{\varepsilon_i -
\varepsilon_a}\right] = 0,
\end{align}
where $\hat K$ is the nonlocal HF exchange potential. Throughout this work $i, j, \ldots$ 
denote occupied orbitals, $a, b, \ldots$
unoccupied orbitals and $p, q, \ldots$  are used for general (i.e. occupied or
unoccupied) orbitals.
The multiplicative exchange--only OEP potential,
$v_{\mathrm{x}}(\mathbf{r})$,
corresponds to the functional derivative of the exchange--only
energy functional, $E_\mathrm{x}[\varphi_{\mathrm{KS}}]$, which has the form
of the
usual exchange--energy functional from HF theory evaluated on KS orbitals:
\begin{align}\label{Eq:ex}
E_\mathrm{x}[\varphi_\mathrm{KS}] = -\frac{1}{2} \sum_{i,j} (ij | ji),
\end{align}
with the two electron integrals defined as
\begin{align}
(pq | rs) = \int \varphi_p^*(\mathbf{r}) \varphi_q(\mathbf{r})
 \frac{1}{|\mathbf{r} - \mathbf{r}'|} \varphi_r^*(\mathbf{r}^\prime)
 \varphi_s(\mathbf{r}^\prime) \mathrm{d} \mathbf{r} \mathrm{d}
 \mathbf{r}^\prime .
\end{align}
The KS orbitals employed are the solutions to the standard KS equation
\begin{align}\label{Eq:KS}
\left[ -\frac{1}{2} \nabla^2 + 
v(\mathbf{r}) + \int
\frac{\rho(\mathbf{r}^\prime)}{|\mathbf{r}-\mathbf{r}^\prime|}
 \mathrm{d} \mathbf{r}^\prime + v_{\mathrm{xc}}[\rho](\mathbf{r})
 \right]
\varphi_p(\mathbf{r}) =
\varepsilon_p \varphi_p(\mathbf{r}),
\end{align}
where $v(\mathbf{r})$ is the external potential due to the nuclei, the
third term is the classical Coulomb potential and $v_{\mathrm{xc}}(\mathbf{r})$
is the local exchange--correlation potential.

Fulfilling the requirement of Eq.~(\ref{Eq:dencon}) through second--order, $\delta \rho^{(2)}(\mathbf{r})=0$, allows the definition of the 
orbital--dependent OEP2 equations for the second--order correlation
potential. In this paper we will use the OEP2--sc \cite{bartlett:2005:abinit2} 
variant of this approach. 
The OEP2--sc correlation functional takes  the standard form of the
 second--order energy expression in many--body perturbation theory [MBPT(2)],
\begin{equation}
E^{(2)}_c=
\frac{1}{2}\sum_{i,j,a,b}\frac{|(ia|jb)|^2}{d_{ijab}}-
\frac{1}{2}\sum_{i,j,a,b}\frac{(ia|jb)(aj|bi)}
{d_{ijab}} 
+\sum_{i,a}\frac{|f_{ia}|^2}{d_{ia}}. \label{EKSS}
\end{equation}
where the denominators are defined as
\begin{equation}
d_{ia}=f_{ii}-f_{aa} \;\;\; \text{and} \;\;\; d_{ijab}=f_{ii}+f_{jj}-f_{aa}-f_{bb},
\end{equation}
where $f_{pq}$ are the usual Fock matrix elements defined in terms of KS--OEP
spin--orbitals,
\begin{align}
f_{pq}=\varepsilon _{p}^{\mathrm{KS}}\delta _{pq}-\langle
p|\hat{K}+v_{\mathrm{xc}}|q\rangle.
\end{align}
For OEP2--sc the semi--canonical (SC) transformation of the OEP2--KS
orbitals is performed \cite{bartlett:1995:coupled},
to reinstate orbital  invariance of the MBPT(2) energy for
rotations which mix occupied or virtual orbitals among themselves. 

This method has been
found to provide a stable alternative to the other second--order correlated 
OEP2--KS theories~\cite{bartlett:2005:abinit2,lotrich:2005:vdw,grabowski:2008:ijqc,grabowski:2010:molphys,grabowski:2011:jcp},
where problems with convergence, overestimation of the correlation energy,
and in many cases poor quality of the correlation potentials were encountered~\cite{engel:2005:oeppt2,mori-sanchez:2005:oeppt2,grabowski:2007:ccpt2,grabowski:2008:ijqc,grabowski:2011:jcp}.
Recently the scaled--opposite--spin version of the second--order correlated
OEP method (SOS--OEP2) was also proposed~\cite{sos_oep2:prb}.

\par In order to solve the OEP equations to determine the 
exchange--correlation  potential, which
is then used in the iterative self--consistent--field (SCF) solution of the KS equations
(see Eq. (\ref{Eq:KS})), we use
the finite basis set implementation of the OEP method from Refs.~\onlinecite{gorling:1999:OEP,ivanov:1999:OEP}.  It involves a projection method
\cite{ivanov:1999:OEP,ivanov:2002:OEP} for solving
the required integral equation, and by construction all potentials are
expanded in terms of auxiliary Gaussian functions.

To minimize 
computational difficulties that are often encountered in the application of
the finite basis set OEP procedure to both  the exchange--only energy
functional~\cite{hirata:2001:OEPU,elp,Hessel1,Joubert1,Kollmar1,Teale1,Glushkov1,Theophilou1}
and that including correlation~\cite{grabowski:2005:1shot,grabowski:2007:ccpt2},
which are  the  manifestation of the well--known
instability associated with numerical solutions of Fredholm integral
equations of the first kind, we  use 
the same  carefully chosen uncontracted basis sets to represent
 both the orbitals and potentials in our 
procedure \cite{hirata:2001:OEPU,ivanov:2002:OEP,bartlett:2005:abinit2,grabowski:2007:ccpt2,grabowski:2008:ijqc}.
These issues have been discussed extensively in the literature and several
schemes have been proposed for managing this
problem~\cite{elp,Hessel1,Joubert1,Kollmar1,Teale1,bulat:2007:sn,Glushkov1,Theophilou1}.
Since we use finite Gaussian--type basis sets, which at
 large $r$ decay much faster than $1/r$,  an incorrect
asymptotic behaviour of our  exchange--correlation  potentials
 (which should decay as $-1/r$) would be obtained.
To ensure the correct asymptotic behaviour we 
use the Colle--Nesbet \cite{Colle:2001:asympt} decomposition of the exchange--correlation OEP  potential into two components,
\begin{equation}
v_{\text xc}(\mathbf{r})=v_{\text{Slater}}(\mathbf{r})+\sum_{t}^{N_{\text{aux}}}
c_t g_t(\mathbf{r}),
\end{equation}
The first term  
\begin{equation}
v_{\text{Slater}}(\mathbf{r}) =
- \sum_{ij}\frac{\phi_i^*(\mathbf{r})\phi_j(\mathbf{r})}{\rho(\mathbf{r})}
\int d\mathbf{r'}
\frac{\phi_j^*(\mathbf{r'})\phi_i(\mathbf{r'})}{|\mathbf{r}-\mathbf{r'}|}+
\mbox{c.c.}
\end{equation}
is the Slater potential  \cite{slater:1951:xpot}, which 
is responsible here for preserving the -1/r asymptotic behaviour and is
calculated on a grid. 
The second component is determined via the OEP integral equation, where 
$g_t(\mathbf{r})$ are the auxiliary Gaussian basis functions and $c_t$ are the
expansion coefficients.

\par
An alternative approach that approximates the exchange--only OEP potential is the
so called localized Hartree--Fock (LHF) method \cite{lhf1,lhfasy}. 
In this method a local exchange potential is derived starting 
from the approximate assumption that the HF and 
the exact--exchange Slater determinants
are equal \cite{lhf1}. The resulting LHF exchange potential is  
\begin{eqnarray}\label{LHF}
&&v_\text{x}^{\text{LHF}}(\mathbf{r}) =  v_{\text{Slater}}(\mathbf{r}) 
+ \sum_{i,j}{'}\frac{\phi_i(\mathbf{r})\phi_j(\mathbf{r})}{\rho(\mathbf{r})}\times\\
\nonumber
&&\ \ \times\int\phi_i(\mathbf{r}')
\left[\sum_k\frac{\phi_k(\mathbf{r})\phi_k(\mathbf{r}')}{|\mathbf{r}-\mathbf{r}'|}-v_{\text{x}}^{\text{LHF}}(\mathbf{r}')\right]
\phi_j(\mathbf{r}')d\mathbf{r}'\ ,
\end{eqnarray}
where the molecular orbitals are assumed to be real.
The second term on the right hand side is the so called correction term and
in its calculation
the HOMO element must be excluded from the double summation (as
indicated by the prime) to enforce the correct asymptotic
behaviour of the LHF potential \cite{lhf1}. If all the $i \neq j$
terms are dropped from the summations in Eq.~(\ref{LHF}), the KLI potential~\cite{krieger:1990:OEP}
is recovered.
We note that, despite the fact that the LHF potential is not a functional 
derivative of any energy functional \cite{staroverov1,staroverov2},
it is computationally very stable and generally regarded as a very good approximation
to the KS exact exchange potential, having been applied to  
a range of different problems in quantum chemistry 
\cite{weimer03,thio,goldlhf,interlhf,ijqc,lhfexci,lhfemb,ipgold}. 
Moreover, it is also derived within 
the  common energy denominator approximation (CEDA) \cite{ceda} and
effective local potential  (ELP) \cite{elp} methods as well as 
the first order approximation to a linear Sham--Schl\"uter equation
\cite{ls2}. However, some studies have noted that the subtle differences between the OEP exchange 
only and LHF/KLI potentials can have significant effects in the calculation of response properties~\cite{Teale:2004:CPL,Gorling:2004:CPL,Teale:2005:PCCP}. 
The quality of the LHF/KLI approximations is further examined in the present work. 

\par To analyze the performance of the orbital--dependent
and standard KS--DFT methods,
 we use as a reference the electronic densities calculated at 
the exchange--only (HF) level, at the
 second--order M{\o}ller--Plesset (MP2) and  coupled--cluster singles--doubles 
with perturbative triples (CCSD(T)) levels.

\subsection{Computational Details}\label{subsec:compdet}
To compare the quality of different WFT and DFT methods, we
have performed calculations for several representative systems, which can be
divided into three classes: 
i) atoms (He, Be, Ne, Ar),  
ii) the neon isoelectronic series ( Si$^{4+}$, Ca$^{10+}$, Zn$^{20+}$), without relativistic corrections, 
and 
iii) small molecules (He$_2$, N$_2$, CO, H$_2$O).

The calculations have been performed using different computational approaches.
In the \textit{ab initio} DFT category we  used
the exchange--only OEPx and correlated OEP2--sc  methods, as implemented 
in  the ACES II package~\cite{aces2}. The
OEP equations were solved in a fully self--consistent manner together with the KS equations until a
final convergence criteria of $10^{-8}$ a.u. on the maximum change in density matrix elements
is reached.

As effective exact--exchange methods we considered the  orbital--dependent LHF and KLI,
 as implemented in the TURBOMOLE program package~\cite{TURBOMOLE}.
 The Slater potential was computed numerically \cite{lhfasy} and the correction term using the conjugate gradient technique\cite{lhf1}.
Hartree-Fock orbitals have been used as starting orbitals for LHF/KLI calculations, 
giving convergence in less than 10 SCF cycles for all systems considered in this work.
The energy and density convergence criteria were set to $10^{-6}$ a.u.

 Among the  standard
\textit{ab initio} WFT methods, MP2 and CCSD(T) calculations have been
 employed to calculate correlation energies and electronic densities. 

The inverse--KS calculations were performed with a development version of the \textsc{Dalton2011} quantum 
chemistry program \cite{DALTON}. The electronic densities for the MP2 and CCSD(T) methods are obtained from relaxed density
 matrices~\cite{hansch:1984:zvec,ricamo:1985:relden,bartlett:1986:relden,bartlett:1989:relden}
constructed using the Lagrangian approach~\cite{helg:1989:lag,jorg:1988:lag,koch:1990:lag,hald:2003:lag}. In order to determine reference KS potentials, eigenvalues and HOMO--LUMO gaps corresponding to the WFT densities we have employed 
the inversion approach of Yang and Wu~\cite{wu:2003:wy}. We employ the same uncontracted basis sets for the expansion of the potential 
and orbitals in this approach. The Fermi--Amaldi potential is used to ensure correct asymptotic decay of the calculated XC potentials.
The smoothing norm procedure of Heaton--Burgess {\emph et al.}~\cite{heaton:2007:sn,bulat:2007:sn} was employed with a regularisation 
parameter of $10^{-5}$ and the calculations were considered converged when the gradient norm was below $10^{-8}$ a.u. For further details see 
Refs.~\onlinecite{wu:2003:wy,grabowski:2011:jcp}.  The HOMO energies and HOMO--LUMO gaps determined from these calculations for the MP2 and CCSD(T) 
densities are labelled KS[MP2] and KS[CCSD(T)], respectively.\\

\subsubsection{Basis-sets}
The selection of the basis sets in this work was mainly dictated by
the requirement of smooth and well-behaved convergence of the OEP calculations.
For this reason all basis sets were constructed by partial or full 
uncontraction of medium size (triple zeta) basis sets originally 
developed for correlated calculations. The choice of the basis functions and
the de--contraction schemes were optimized to ensure a 
smooth behaviour of OEP potentials, especially because
in all calculations where the KS potentials needs to be expanded in terms 
of Gaussian basis functions the same basis sets was employed for the 
potential expansion as for the molecular orbitals.
 
In more detail, an even tempered 20s10p2d basis was employed 
for He atom and He$_2$ molecule; 
the uncontracted ROOS--ATZP~\cite{widm90} basis was used for Ne;  
for the Be atom the ROOS--ATZP basis set was used with s and p 
functions uncontracted; 
for Ar the uncontracted ROOS--ATZP \cite{widm90} basis is used for 
s and p type basis functions, whereas for $d$ and $f$ orbitals 
we used the uncontracted aug-cc-pwCVQZ \cite{peterson02} basis set. 
The uncontracted cc--pVTZ
basis set of Dunning~\cite{dunning:1989:bas} was used for the 
molecular systems N$_2$, CO and H$_2$O.
For the neon isoelectronic series members the following basis sets were used: 
in case of Si$^{4+}$ the ROOS--ATZP basis set with s and p 
functions uncontracted;
for Ca$^{10+}$ ion we used uncontracted ROOS-ATZP basis set of Ne; 
the Zn$^{20+}$ ion was calculated in the ROOS-ATZP basis set with s 
functions uncontracted and the g functions removed.

We remark that with this choice of the basis sets, especially for ionic systems,
our HF and CCSD(T) results differ from benchmark results
\cite{davids:1991:exact,ver}.
Nevertheless, because the main goal of the present work is to perform a 
relative (and mostly qualitative) comparison between different 
methods and because all 
the exchange-only as well as all the XC methods considered here have 
a similar basis set convergence behaviour (almost linear for exchange and cubic 
for correlation), the analysis of the different results is expected
to be only slightly influenced by this issue. Thus, the present
computational set up should allow fair comparison and assessment of
the different approaches. 

\section{Results}

In this section we compare a range of orbital--dependent 
exchange--only and exchange--correlation functionals with reference 
results from WFT methods.
Different criteria are used to assess the quality of the approaches. 
Firstly in Section~\ref{subsec:energy}, we assess the accuracy of the 
total energies delivered by each method. 
Then we consider in Section~\ref{subsec:dens} the density differences (DRDs for atomic systems) 
relative to Hartree-Fock to assess both 
the impact of correlation and the effect of approximations in the derivation of KS-potentials 
on the electronic density obtained. 
Finally, in Section~\ref{subsec:HL} 
we compare the HOMO-LUMO gaps and HOMO energies calculated for each approach.

\subsection{Total energies}
\label{subsec:energy}
The total ground state energies are presented in Table \ref{totenetab}.
For each method the mean absolute error (MAE) 
with respect to the reference values (Hartree-Fock and CCSD(T) results 
for exchange-only and exchange-correlation methods, respectively) are also reported. 
The MAEs are separated for each class of systems 
(atoms, neon isoelectronic series, molecules). 
The total MAE is also reported in the last line of  Table \ref{totenetab}.
\begin{table*}
\begin{center}
\caption{\label{totenetab} Total energies (in Hartree) for Hartree-Fock and CCSD(T) methods. Differences (in mHartree) from these reference values are shown for several exchange-only (X-only) and exchange-correlation methods, respectively. Mean absolute errors (MAE)  
are reported for all systems as well as for the atomic, neon isoelectronic series and the molecular systems.}
\begin{ruledtabular}
\begin{tabular}{lrrrrcrrr}
 & \multicolumn{4}{c}{X-only methods} & & \multicolumn{3}{c}{XC methods} \\
\cline{2-5}\cline{7-9}
System & HF &	OEPx &	LHF &	KLI & $\;\;$ &	CCSD(T) & MP2 &	OEP2-sc \\
\hline
He	& -2.8617      & 0.0 & 0.0 & 0.0 & &	-2.9025 & 5.6 &	5.6 \\
Be	& -14.5730     & 0.6 & 0.6 & 0.7 & & -14.6619 & 19.9 & 19.7 \\
Ne	& -128.5466    & 1.6 & 2.3 & 2.2 && -128.9000 & 7.5 & 5.9 \\
Si$^{4+}$& -285.1801  & 1.2 & 7.7 & 7.7 && -285.3462 & 2.4 & 2.1 \\
Ca$^{10+}$& -640.3120 & 1.9 & 3.1 & 3.3 && -640.5276 & 3.1 & 3.0 \\
Zn$^{20+}$& -1552.5576 & 0.8 & 2.5 & 2.9 && -1552.7104 & 1.4 & 1.4 \\
Ar	& -526.8165     & 5.3 & 7.1 & 7.2 && -527.4575 & 23.9 & 22.0 \\
He$_2$  & -5.7234       & 0.0 & 0.0 & 0.0 && -5.8051 & 11.1 & 11.1 \\
N$_2$   & -108.9847    & 5.3 & 7.5 & 7.7 && -109.4763 & 20.1 & 11.6 \\
CO	& -112.7816     & 5.2 & 7.2 & 7.4 && -113.2574 & 23.5 & 14.5 \\
H$_2$O	& -76.0578      & 2.0 & 3.2 & 3.5 &&	-76.3869 & 14.9 & 12.7 \\
\hline
MAE$_{\text{ato}}$ & &	1.9 &	2.3&	2.5&&&		14.2&	13.3\\
MAE$_{\text{iso}}$ &&	1.4 &	3.6&	4.0&&&		3.6&	3.1\\
MAE$_{\text{mol}}$ &&	3.1 &	4.5&	4.6&&&		17.4&	12.5\\
\hline
MAE$_{\text{tot}}$ && 	2.2 &	3.7&	3.9&&&		12.1&	10.0\\
\end{tabular}
\end{ruledtabular}
\end{center}
\end{table*}

Considering exchange-only methods, we see that they all perform very 
similarly and are extremely close to
the reference HF results. This finding indicates the high quality of these
approaches for the description of energetic properties of electronic systems.
In closer detail, we observe that the OEPx
method delivers the  smallest deviations from HF, as expected.
Instead, because of the variational principle, LHF and KLI results are
systematically slightly above the HF energy 
(except for case of He, which gives exactly the same energy, by definition). 
Nevertheless, the LHF and KLI approximations show reasonable agreement with the OEPx results for the 
neutral atomic and molecular systems, typical differences 
between HF and LHF/KLI being below 8 mH. 
Overall, the results of Table~\ref{totenetab} show that LHF and KLI have a fairly similar performance,
with error measures approximately twice as large as OEPx. Moreover, the deviations obtained with LHF
are slightly smaller than the ones yielded by KLI.  
In this respect, it should be noted 
that both methods employ the same (i.e. Hartree-Fock) total energy expression but different
(non-variational) potentials. Thus, the difference between LHF and KLI originates only from self-consistent effects.
In addition, it is worth noting that the KLI potential is not 
invariant with respect to orbital rotations, whereas the LHF energy is stable in this respect
\cite{lhf1}.

Considering the full exchange-correlation approaches we can see that the OEP2-sc method reproduces 
MP2 results quite well and both have small deviations with respect  to CCSD(T), with a total MAE of about 10-12 mH.
This behaviour can be attributed to the fact that OEP2-sc has a second--order correlation potential 
based on the MBPT(2) type energy functional.
Interestingly, the OEP2-sc MAEs are also slightly lower than the MP2 ones, which may be a result of the correlation
effects included in the orbitals by relaxation during the self--consistent solution of the KS-OEP equations. 
Examining the differences  between the errors for different classes of systems, it is interesting to note 
that the neon isoelectronic series  gives MAEs 4-5 times smaller than the ones 
for the atomic and molecular systems, for both the MP2 and OEP2-sc methods.
This result can be understood from the relatively simple behaviour of the correlation energies for 
the heavier ions~\cite{davids:1991:exact,ver}, which are very well described by
second--order perturbation theory.

\subsection{Density-based Analysis}\label{subsec:dens}

For a more detailed comparison about the potential of orbital--dependent methods, we directly compare the
DRD distributions calculated relative to the HF radial densities for the atomic systems. 
This allows us to inspect the influence of correlation effects on 
the density. 

\begin{figure}
\center
\includegraphics[width=\columnwidth]{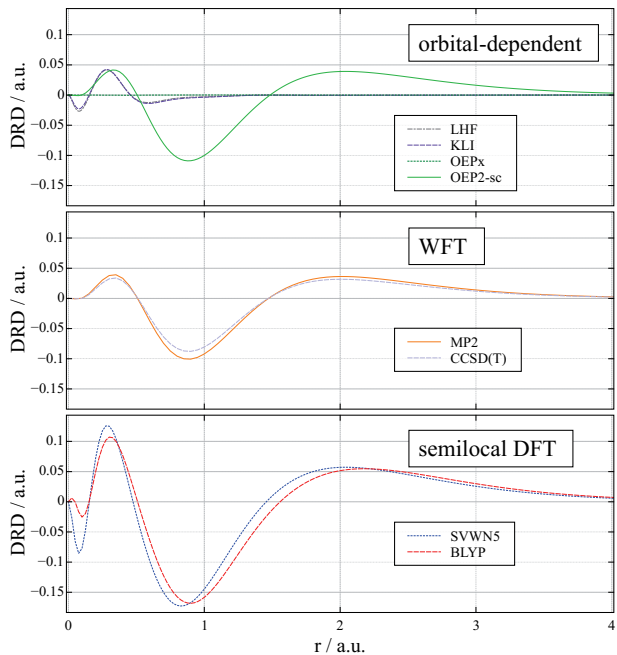}
\caption{ \label{fig_ne_drd} Difference radial density distributions (DRD; see Eq. (\ref{drd_eq}))
for the neon atom, calculated using 
orbital--dependent OEP--DFT functionals (top panel)
 wave function theory (WFT) methods (middle panel)
and semi-local standard DFT methods (bottom panel).}
\end{figure}

In Figure~\ref{fig_ne_drd} we compare DRDs for the neon atom, calculated
using orbital--dependent KS functionals (OEPx, KLI, LHF and 
OEP2-sc; top panel) and wave function theory methods (MP2 and CCSD(T); 
middle panel).
We also report, for comparison, (semi-)local standard DFT methods 
(SVWN5 \cite{slat74,vosko:1980:VWN}, BLYP\cite{lee:1988:BLYP,becke:1988:GGA};
 bottom panel). The CCSD(T) plots are our reference results. 

Examining the plots we note that the 
OEPx DRD is flat and almost overlaps with the $x$--axis (the
OEPx density is almost identical to HF one).
Including correlation in the orbital--dependent calculations at the OEP2-sc 
level provides a DRD that closely resembles the
reference MP2 and CCSD(T) results.
The SVWN5 and BLYP results in Figure \ref{fig_ne_drd}  
show a similar behaviour to the
reference CCSD(T) DRD, except for their amplitudes, which are much larger
than the CCSD(T) results. However, this reasonable overall behaviour arises because 
dynamic correlation effects are represented mainly by the exchange-only S and B88
functionals, which dominate DRDs obtained from the total exchange-correlation
SVWN5 and BLYP calculations respectively
\cite{Jankowski:2009:DRD,Jankowski:2010:DRD}.

Interestingly, the LHF and KLI DRDs in Figure \ref{fig_ne_drd} are 
not as flat as may be expected based  on the analysis of their 
total energies as presented in Section~\ref{subsec:energy}. 
It appears that for $r < 1.$ a.u.  the LHF/KLI approximations 
lead to substantial differences in the DRDs relative to HF but that 
their accuracy improves rapidly as $r$ increases. Furthermore, 
these errors give rise to a
DRD profile that to some extent mimics correlation effects at small $r$. 
Examining the top panel in Figure~\ref{fig_ne_drd} we see that 
first downward peak (at $r\approx 0.08$ a.u.) is also present in the SVWN5 result.
This peak corresponds to a large density deviation at the nucleus (i.e. LHF/KLI has less density at the nucleus than HF): 
however due to the radial factor in Eq. (\ref{drd_eq}) it appears as a small peak
at finite $r$ (see also section \ref{sec:mol}).
The following upward peak (at $r\approx 0.28$ a.u.) in the inter-shell region closely resembles 
the one of OEP-sc, MP2 and reference CCSD(T). 
The outer features of the correlated DRDs are however not 
mimicked at the KLI/LHF level. Similar observations were
made by Teale and Tozer in Ref.~\onlinecite{Teale:2005:PCCP} and used to 
interpret the fact that LHF and KLI approaches yielded unexpectedly 
accurate NMR shielding constants in Refs.~\onlinecite{Teale:2004:CPL} 
and~\cite{Gorling:2004:CPL}.
Whilst  energetically the LHF and KLI approximations are reasonably accurate their potentials are not the functional derivative
of the orbital--dependent exchange energy with respect to the density. Instead errors (relative to OEPx) associated with the approximations 
used in their derivation lead to potentials that give rise to densities with errors at small $r$ in the core and near valence regions, 
as manifested in the DRDs. The results here may go some way to further explaining the results in Refs.~\onlinecite{Teale:2004:CPL} and~\cite{Gorling:2004:CPL} 
for NMR parameters, since these properties are sensitive only to regions close to nuclei and in these areas the LHF and KLI results mimic correlation 
effects. This may explain why the calculated values exhibit a quality closer to correlated GGA results than to OEPx.\\    

\subsubsection{Neon isoelectronic series}
\begin{figure}
\center
\includegraphics[width=\columnwidth]{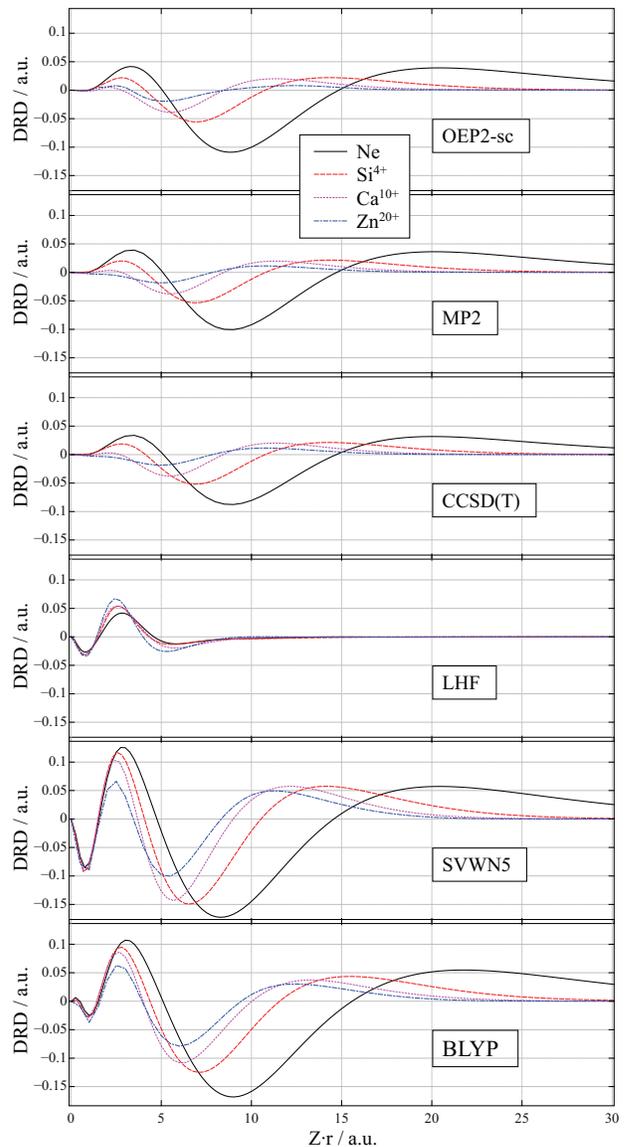}
\caption{ \label{fig_iso_drd_scaled} Difference radial density distributions (DRD; see 
Eq. (\ref{drd_eq})) for the Ne atom and a few members of the neon
isoelectronic series (Si$^{4+}$, Ca$^{10+}$ and Zn$^{20+}$), calculated
using 
orbital--dependent OEP--DFT functionals (OEP2-sc, KLI and LHF),
 wave function theory methods (MP2 and CCSD(T))
and semi-local standard DFT methods (SVWN5, BLYP). Note that on the $x$-axes the radial coordinate $r$ has been scaled by the nuclear charge,
$Z$.}
\end{figure}

To analyse the behaviour of the different approximations in more detail, we present 
similar plots for a few members of the neon
isoelectronic series i.e.  Si$^{4+}$, Ca$^{10+}$, Zn$^{20+}$ and Ne atom in 
Figure \ref{fig_iso_drd_scaled} . For clarity we present results obtained
from different methods in separate panels. Moreover, we do 
not report KLI and OEPx results here, because they are 
essentially indistinguishable 
from LHF and HF results, respectively, on the scale presented.
We observe that a clear trend can be distinguished for CCSD(T) and MP2.
The DRDs become, of course, more compact with increasing $Z$ values 
(note that in the plot $Zr$ is reported on the $x$-axis).
At the same time the height of the different peaks is reduced with $Z$.
This trend is more accentuated for core features that are almost invisible (on 
this scale) already for Ca$^{10+}$.
The reference trend is well reproduced by the OEP2-sc calculations.
Thus, this method proves to describe the change in correlation effects on 
increasing $Z$ with good accuracy and
reliability.

The same trend is also qualitatively reproduced by pure DFT functionals.
However, these approaches fail to give a correct quantitative description
of the DRDs of the different members of the isoelectronic series: 
i) the amplitude of the oscillations is overestimated , 
ii) the decrease of the peak height with Z is slower 
and iii) an additional
downward peak is observed near the origin.

In contrast, the LHF calculations display a completely different
behaviour. In this case, in fact, the DRD peak position is almost fixed 
at $Z r\approx 2.5-2.8$ a.u.  and the
amplitude of the DRD oscillations is almost constant (actually slightly increasing with increasing values of $Z$). 
This is opposite to what can be expected on the
basis of the accurate CCSD(T) calculations and so the effect of LHF mimicking correlation breaks 
down as $Z$ increases.
These results show therefore that the LHF potential does not in general include
proper correlation effects, as expected from an orbital--dependent exchange-only method.
 Rather, the features of the DRD plots
can be traced back to some limitations of the correction term
which cannot mimic accurately the exact-exchange response term in the 1s-2s inter-shell region.
On the other hand, the fact that the DRD profiles are almost independent from $Z$ and vanish in the valence region 
means that the LHF potential deviation from the OEPx potential is only due to an almost constant term near the core region.
This error is therefore quite systematic and LHF can be safely used as a reasonable
 approximation to investigate exact-exchange in heavy ions.\\

\subsubsection{Molecules}
\label{sec:mol}
\begin{figure}
\center
\includegraphics[width=\columnwidth]{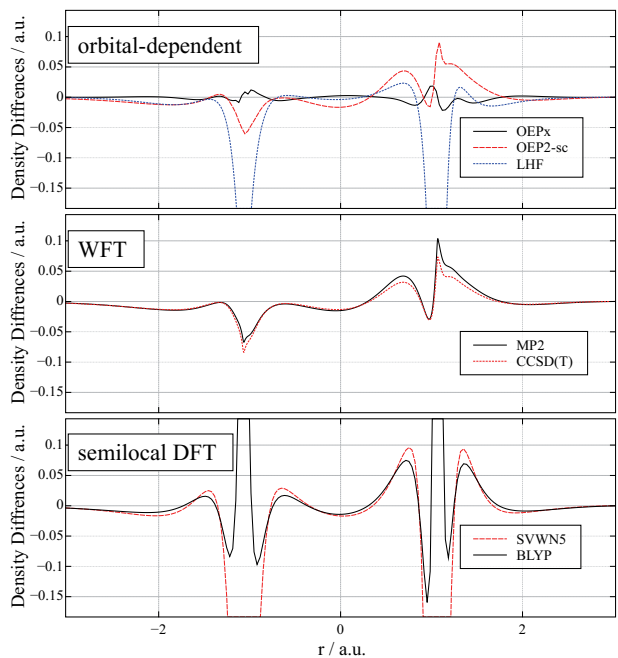}
\caption{ \label{fig_co_drd} Difference total density distributions 
(along the molecular axis) relative to HF, for the CO molecule, calculated
using 
orbital--dependent OEP--DFT functionals (top panel)
 wave function theory methods (middle panel)
and semi-local standard DFT methods (bottom panel).}
\end{figure}

The density-based analysis presented so far for atoms using DRDs, can 
be carried out also for molecules by considering density differences just along one line. 
To show this we report in Figure~\ref{fig_co_drd}
the density difference along the molecular axis of a CO molecule, 
relative to HF, for different theoretical approaches.
The shape of the accurate MP2 and CCSD(T) density differences reflects 
a depletion of the density around the carbon atom and an increase 
around the oxygen atom relative to HF. This in turn reflects the fact that HF 
theory predicts a qualitatively incorrect dipole moment for CO
($-$0.104 a.u. for HF compared with $+$0.111 a.u. for MP2),
which is corrected in the correlated methods by a redistribution of charge. 
We see once again that the OEP2-sc method 
can reproduce the reference CCSD(T)
and MP2 results with good accuracy. 
In contrast, both 
LHF and the conventional DFT functionals
show large  density-difference peaks around nuclei 
which are however also present in atomic systems, see Section
\ref{subsec:dens}.
The LHF density-difference profile has qualitatively more in common with
SVWN5 whereas in BLYP the peaks near the nuclei are of opposite sign.  
This is consistent with the analysis in Ref.~\onlinecite{Teale:2005:PCCP}.

\subsection{KS HOMO-LUMO gaps and  Ionization Potentials}\label{subsec:HL}
To complete our analysis we consider the KS HOMO--LUMO gaps and HOMO energies delivered by each of the methods
in comparison with accurate values obtained from KS[CCSD(T)] calculations. 
The latter have been used here as reference values for the orbital energies and energy-gaps
to allow a direct comparison between all of the methods using the same finite Gaussian basis sets. 
For many of the systems considered here benchmark estimates of the orbital energies may 
be found in
Refs.~\onlinecite{Petersilkagrosburke:2000:HL,magyar:2004:HL,wilsontozer:2002:HL,radzig:1985:IP,grabo:1995:IP,last3}. However, 
these values are calculated using a range of different basis sets and different methodologies 
to obtain accurate densities, making consistent comparisons with our data difficult. Nonetheless we note 
that our values are broadly consistent with those in
Refs.~\onlinecite{Petersilkagrosburke:2000:HL,magyar:2004:HL,wilsontozer:2002:HL,radzig:1985:IP,grabo:1995:IP,last3}.

The KS HOMO--LUMO 
gaps for each approach are presented in Table~\ref{table3}.
\begin{table*}
\begin{center}
\caption{\label{table3}
HOMO--LUMO gaps (in eV) from various methods (see text for further details).
Mean absolute relative errors (MARE) with respect to KS[CCSD(T)] values are presented for the full set of systems as well as for the subsets corresponding to neutral atoms (ato), the members of the neon isoelectronic series (iso) and molecular systems (mol). 
} 
\begin{ruledtabular}
\begin{tabular}{lrrrrrr} 
System & OEPx & OEP2-sc & KLI & LHF & KS[MP2] & KS[CCSD(T)] \\ 
\hline
He & 21.60 & 21.32 & 21.67 & 21.67 & 21.33 & 21.21 \\ 
Be & 3.57 & 3.64 & 3.54 & 3.54 & 3.63 & 3.61 \\ 
Ne & 18.48 & 16.45 & 18.47 & 18.48 & 16.81 & 17.00  \\ 
Si$^{4+}$ & 104.02 & 100.61 &  105.11&	105.11  & 102.37& 102.45 \\ 
Ca$^{10+}$ & 348.35 & 346.03 &  349.43&	349.43  & 346.13	&346.17  \\ 
Zn$^{20+}$ & 1056.56 & 1046.56 &  1055.78&	1055.76  &  1052.45&	1052.43  \\ 
Ar & 11.41 & 11.43 & 11.06 & 11.09 & 11.45 & 11.51 \\ 
He$_2$ & 21.28 & 21.02 & 20.82 & 20.82 & 20.43 & 20.31 \\ 
N$_2$ & 9.21 & 8.37 & 8.73 & 8.78 & 8.41 & 8.51 \\ 
CO & 7.77 & 7.22 & 7.32 & 7.31 & 7.27 & 7.25 \\ 
H$_2$O & 8.47 & 7.51 & 8.52 & 8.32 & 7.55 & 7.71  \\  
\hline
MARE$_{\text{ato}}$ & 3.1\% & 1.3\% & 4.2\% & 4.1\% & 0.7\% \\
MARE$_{\text{iso}}$ & 0.9\% & 0.8\% & 1.3\% & 1.3\% & 0.0\% \\
MARE$_{\text{mol}}$ & 7.5\% & 2.0\% & 4.1\% & 3.6\% & 1.0\% \\
\hline
MARE$_{\text{tot}}$ & 4.1\% & 1.4\% & 3.4\% & 3.2\% & 0.6\% \\
\end{tabular}\\
\end{ruledtabular}
\end{center}
\end{table*}
The exchange-only methods show in general an overestimation with respect
to the KS[CCSD(T)] values. However, for argon and beryllium the opposite trend
is obtained. This shows that the correlation effects are subtle 
in this context and cannot be easily predicted.
For the atomic and ionic systems the LHF and KLI gaps are rather 
close to the OEPx ones. For molecules the gaps are 
reduced compared to OEPx and so move closer to the KS[CCSD(T)] values, 
leading to an overall reduction in the error measures: 
the mean absolute relative error for molecules (MARE$_{\text{mol}}$) is reduced from 7.5\% in OEPx
to 3.6\% in LHF.

The OEP2-sc method leads to a substantial improvement of the HOMO-LUMO gaps, thanks to
the inclusion of correlation. In particular it yields smaller
gaps than OEPx for all the systems except beryllium and argon, in line with the reference values. Thus,
it appears to be able to provide a qualitatively correct description of
correlation effects in all systems, unlike conventional DFT correlation 
functionals which always increase the HOMO-LUMO gap \cite{ls2}. 
However, this result may benefit from a strong cancellation of 
systematic errors as indicated by  the analysis of HOMO energies (see below).
Moreover, the OEP2-sc correlated approach leads for many systems to
a too strong reduction of the gap, resulting in an 
underestimation of the KS[CCSD(T)] values. 
This makes it somewhat further from the KS[CCSD(T)] 
values than KS[MP2]. This finding supports the idea
that OEP2-sc results for the HOMO--LUMO gap should be treated with caution
and may also indicate 
that the orbital relaxation effects incorporated in OEP2-sc have 
a significant effect on determining the Kohn--Sham 
eigenvalue spectrum. 
Nevertheless, overall OEP2-sc is the best of the DFT approaches 
considered in the present work and is substantially more accurate than typical conventional DFT functionals.

\par In Table \ref{table4} we present the HOMO energies 
together with their MAREs, 
obtained with the same methods as in Table \ref{table3}.
\begin{table*}
\begin{center}
\caption{\label{table4}
HOMO energies (in eV) obtained from various methods (see text for further details).
Mean absolute relative errors with respect to the KS[CCSD(T)] values are also presented for the full set of systems as well as for the subsets corresponding to neutral atoms (ato), the members of the neon isoelectronic series (iso) and molecular systems (mol).}
\begin{ruledtabular}
\begin{tabular}{lrrrrrrr}
~ & OEPx & OEP2-sc & KLI & LHF & HF & KS[MP2] & KS[CCSD(T)] \\ 
\hline
He                           & -24.98 & -24.70 &  -24.99 & -24.99 & -24.99 & -24.68 & -24.56  \\ 
Be                           & -8.41 & -8.66 & -8.34 & -8.41 & -8.42 & -9.07 & -9.43  \\ 
Ne                           & -23.15 & -20.97 & -23.11 & -23.12 & -23.14 & -21.01 & -21.14  \\ 
\multicolumn{1}{l}{Si$^{4+}$} & -168.53 & -161.40 &  -168.46 & -168.47 & -168.52 & -164.08 & -164.16  \\  
\multicolumn{1}{l}{Ca$^{10+}$} & -593.40 & -590.37 &  -593.33 & -593.33 & -593.38 & -586.49 & -586.51  \\ 
\multicolumn{1}{l}{Zn$^{20+}$} & -1847.95 & -1823.08 &  -1843.90 & -1843.88 & -1847.93 & -1826.46 & -1826.30 \\ 
Ar                            & -16.07 & -15.66 &  -16.03 & -16.06 & -16.08 & -15.47 & -15.58  \\ 
He$_2$                       & -24.98 & -24.70 &  -24.83 & -24.83 & -24.98 & -22.74 & -22.62  \\ 
N$_2$                       & -17.16 & -18.30 &  -17.12 & -17.08 & -16.67 & -14.93 & -13.96  \\ 
CO                          & -15.02 & -14.65 &  -15.01 & -14.97 & -15.07 & -13.56 & -13.17 \\ 
H$_2$O                       & -13.70 & -14.25 & -13.58 & -13.66 & -13.75 & -11.44 & -11.53 \\ 
\hline
MARE$_{\text{ato}}$ & 6.3\% & 2.5\% & 6.4\% & 6.2\% & 6.3\% &
1.4\% &  \\ 
MARE$_{\text{iso}}$ & 1.7\% & 0.8\% & 1.6\% & 1.6\% & 1.7\% & 0.0\%
&   \\ 
MARE$_{\text{mol}}$ & 16.6\% & 18.8\% & 16.0\% & 16.1\% & 15.9\%
& 2.8\% &   \\ 
\hline
MARE$_{\text{tot}}$ & 8.8\% & 8.0\% &  8.6\% & 8.5\% & 8.5\% &
1.5\% &   \\ 
\end{tabular}
\end{ruledtabular}
\end{center}
\end{table*}
In this case, all the exchange--only methods (HF, LHF, KLI, OEPx) 
perform rather well compared with KS[CCSD(T)]
and with similar accuracy (MARE$_{\text{tot}}$ below 9\%).
The KLI and LHF HOMO energies 
are very close to each other and slightly closer to the KS[CCSD(T)] values, whereas OEPx
is slightly more similar to HF, consistent with the results in the previous sections. 
On the other hand, the correlated OEP method (OEP2-sc), 
despite being the best DFT approach,
provides only slightly more accurate results relative to KS[CCSD(T)] than OEPx. 
In fact, its MARE$_{\text{tot}}$ is 8\% and thus much larger than for  
KS[MP2] (MARE$_{\text{tot}}$=1.5\%).
Considering the molecular results, the MARE$_{\text{mol}}$ value is much larger for OEP2-sc than for KS[MP2] and the individual 
values are substantially different to the KS[CCSD(T)] reference values. 
Orbital energies are particularly sensitive to the quality of the underlying exchange-correlation potentials and 
these results indicate that the OEP2-sc potential may be improved in this respect. This is particularly clear when 
comparing the OEP2-sc and KS[MP2] HOMO eigenvalues. The densities for these two approaches are typically 
very similar (see Figures~\ref{fig_ne_drd}-~\ref{fig_co_drd}), whilst their HOMO eigenvalues differ substantially. 
For N$_2$ and H$_2$O the effect of OEP2-sc correlation is to make the HOMO eigenvalue more 
negative than OEPx, whilst the reference values are more positive. Given the sensitivity of the HOMO eigenvalue 
to the exchange--correlation potential this type of comparison may be a useful further test of other \textit{ab initio} DFT functionals.

\section{Conclusions}
In this work we have presented a density-difference based analysis of
orbital--dependent exchange 
and exchange--correlation functionals in KS-DFT. The use of \textit{ab initio} DFT methods via the OEP approach gives 
a partitioning of exchange and correlation contributions much more in line with standard \textit{ab initio} WFT methods. 
Comparison of OEPx and HF densities showed that exact-exchange-only DFT densities are very similar to those obtained from HF. 
This is also reflected in the comparison of their energies and HOMO eigenvalues. We also considered the LHF and KLI approximations to OEPx. 
These approaches deliver exchange and total energies that are close to that of OEPx and HF, however, examination of DRDs relative to
 HF revealed substantial differences between the LHF/KLI and HF densities. 
In particular, their densities differ in the core and inner-valence regions close to nuclei.
Although they remain accurate in the outer valence and asymptotic regions. 

To assess the accuracy of exchange--correlation functionals the DRDs associated with MP2 and CCSD(T) 
theories were constructed and compared with standard DFT results and the correlated OEP2-sc approach 
for the neon atom and three members of the neon isoelectronic series.
 The standard DFT distributions showed a reasonable qualitative reproduction 
of the CCSD(T) DRDs  though their 
amplitudes were not highly accurate. The OEP2-sc method delivers results of good accuracy very close to the MP2 DRDs,
 as may be expected. Interestingly, for neon the LHF and KLI DRDs show features close to the nucleus that appear to mimic 
correlation effects in that region. To examine this further the Si$^{4+}$, Ca$^{10+}$ and Zn$^{20+}$ members of the 
neon isoelectronic series were investigated. Here similar conclusions were obtained for the standard DFT functionals which
 qualitatively reproduce the MP2 and CCSD(T) DRDs and also for the OEP2-sc method which closely reproduces the WFT DRDs 
as $Z$ increases. However, for LHF and KLI as $Z$ increases the qualitative behaviour of the DRDs is different, exhibiting 
peaks close to the nucleus that are almost unchanged with $Z$. 
This indicates that the mimicking of correlation in LHF and KLI is 
not a general feature but rather derives from systematic errors of the
correction term in the inter-shell region.

To investigate further, density differences for the CO molecule were considered. 
Again OEPx was found to give density differences close to HF and OEP2-sc gave density differences close to those from 
WFT methods. However, LHF/KLI were found to have a different behaviour
near the nuclei, in agreement with the analysis of the isoelectronic
series and the one in  Ref.~\onlinecite{Teale:2005:PCCP}. Accidentally,
this behaviour is slightly similar to that of the conventional exchange--correlation 
functionals and may go someway towards explaining the observations 
for NMR shielding calculations in Refs.~\onlinecite{Teale:2004:CPL,Gorling:2004:CPL}. 

Finally the HOMO--LUMO gaps and HOMO energies were considered for each of the methods. 
The results were compared  with values calculated corresponding to accurate CCSD(T) electronic densities via the approach of Ref.~\onlinecite{wu:2003:wy}. 
These quantities were found to be a sensitive probe of the quality of the approaches. 
In general, the analysis revealed that correlation effects are quite complex
for these properties, giving different trends for different systems,
and that an accurate description of correlation contributions 
is important for accurate results.
In particular, the study of the HOMO energies indicated that even
at the OEP2-sc level there are evident limitations in the description of 
the correlation potential, so that relatively poor improvements with respect
to the OEPx results can be achieved. This could be partially connected to
the finite basis set implementation of the correlated OEP procedure, in
which the choice of basis set used for the calculations play a crucial role
in the description of subtle correlation effects visible in HOMO energies.
Nevertheless, these limitations
may be thought to be mainly systematic and are thus often 
hidden by error cancellation effects, as in the case of HOMO--LUMO gaps. 

Overall, our results show that the \textit{ab initio} DFT OEPx and OEP2-sc approaches provide 
density-functional exchange and correlation energies that are similar to those in traditional wave 
function approaches. We have also shown that care must be taken when applying the LHF and KLI approximate 
exchange approaches. 
Whilst these approximations may be accurate in terms of their energies they
can
show important differences in the densities and HOMO--LUMO gaps they produce.
Nevertheless it has been recently shown that LHF KS orbital energies yield
very accurate TD-DFT excitation energies
for a wide class of molecular systems \cite{lhfexci}.
Finally, good consistency between the  OEP2-sc and KS[CCSD(T)] results was observed, supporting the idea 
that this method can be  used as a benchmark against which to 
test new density functional approximations. Although, a deeper analysis 
of the HOMO eigenvalues from this approach indicated that some limitations
exist also for this advanced method and further work will be needed
to improve the description of subtle correlation effects.

\section*{Acknowledgments}
It is a pleasure to dedicate this work to Prof. Rodney J. Bartlett on the occasion of
his 70th birthday. We would like to thank Rod for sharing his ideas, many stimulating discussions, and for giving
two of us (I.G, Sz.S) the opportunity to work together at the  Quantum Theory
Project. 

This work was partially supported by the Polish Committee for Scientific
Research MNiSW under Grant no. N N204 560839, the National Science
Center under Grant No.~DEC-2012/05/N/ST4/02079, and the European Research Council (ERC) 
Starting Grant FP7 Project DEDOM, Grant No. 207441. 
A. M. T. gratefully acknowledges support via the Royal Society University Research Fellowship scheme.
We thank TURBOMOLE GmbH for providing the TURBOMOLE program package and M. 
Margarito for technical support.

\end{document}